\documentclass[11pt]{article}
\usepackage[normalem]{ulem}
\pdfoutput=1 

\usepackage{jheppub} 
\usepackage{url}
\usepackage{caption}
\usepackage{subcaption}
\usepackage{extarrows}
\usepackage{lmodern}
\usepackage{microtype}

\usepackage[latin9]{inputenc}
\usepackage{float}
\usepackage{amsmath}
\usepackage{amssymb}
\usepackage{graphicx}
\usepackage{esint}
\usepackage{booktabs,array,enumitem}
\usepackage{hyperref}
\usepackage{comment}
\usepackage{color}
\usepackage{microtype}
\usepackage{cleveref}
\usepackage{breakurl}
\usepackage{bbm}
\newcommand{\be}{\begin{equation}}
\newcommand{\ee}{\end{equation}}
\newcommand{\ben}{\begin{displaymath}}
\newcommand{\een}{\end{displaymath}}
\newcommand{\bea}{\begin{eqnarray}}
\newcommand{\eea}{\end{eqnarray}}
\def\K{K{\"a}hler }
   \newcommand{\rf}[1]{(\ref{#1})}
\newcommand{\vp}{\varphi}

\usepackage{bm}
\usepackage{xspace}

\def\be{\begin{equation}}
\def\ee{\end{equation}}
\def\bea{\begin{eqnarray}}
\def\eea{\end{eqnarray}}
\def\ba{\begin{array}}
\def\ea{\end{array}}
\def\bit{\begin{itemize}}
\def\eit{\end{itemize}}

\def\a{\alpha}

\def\vp{\varphi}

\newcommand{\ld}{\Lambda{\rm CDM}}

\newcommand{\wa}{w_0w_a{\rm CDM}}
\newcommand{\bao}{{\rm BAO}}
\newcommand{\cmb}{{\rm CMB}}

\newcommand{\cmbspa}{\text{CMB-SPA}}
\newcommand{\sn}{{\rm SN}}
\newcommand{\Omk}{\Omega_{k}}
\newcommand{\OmK}{o\Lambda{\rm CDM}}
\newcommand{\owa}{ow_0w_a{\rm CDM}}

\newcommand*{\polybin}{\selectfont\textsc{PolyBin3D}\xspace}

 \makeatletter

\allowdisplaybreaks

\makeatother

\definecolor{blue(pigment)}{rgb}{0.2, 0.2, 0.6}

\makeatletter
\DeclareRobustCommand{\rcite}[1]{%
  \rcite@aux#1,\@nil{#1}%
}
\def\rcite@aux#1,#2\@nil#3{%
  \if\relax#2\relax
    Ref.~\cite{#3}%
  \else
    Refs.~\cite{#3}%
  \fi
}
\makeatother

\hypersetup{
    colorlinks = true,
    citecolor = {blue},
    linkcolor = {blue},
    urlcolor = {blue},
}

\def\be{\begin{equation}}
\def\ee{\end{equation}}

\def\rcite#1{ref.~\cite{#1}}

\title{\rm { \LARGE \bf    Inflation, Open Universes, and Dark Energy}}

\author[a]{Anton Chudaykin,}
\author[b,c]{Mikhail M. Ivanov,}
\author[d]{Renata Kallosh,}
\author[d]{Andrei Linde,}
\author[d,e]{Oliver~H.\,E.\,Philcox,}
\author[f]{and Yusuke Yamada}

\affiliation[a]{D\'epartement de Physique Th\'eorique and Center for Astroparticle Physics,\\
Universit\'e de Gen\`eve, 24 quai Ernest  Ansermet, 1211 Gen\`eve 4, Switzerland}
\affiliation[b]{Center for Theoretical Physics -- a Leinweber Institute, Massachusetts Institute of Technology, Cambridge, MA 02139, USA} 
\affiliation[c]{The NSF AI Institute for Artificial Intelligence and Fundamental Interactions, Cambridge, MA 02139, USA}
\affiliation[d]{Leinweber Institute for Theoretical Physics at Stanford, 382 Via Pueblo, Stanford, CA 94305, USA}
\affiliation[e]{Kavli Institute for Particle Astrophysics and Cosmology, 382 Via Pueblo, Stanford, CA 94305, USA}
\affiliation[f]{Cosmology, Gravity, and Astroparticle Physics Group, Center for Theoretical Physics of the Universe,
Institute for Basic Science (IBS), Daejeon, 34126, Korea}
\emailAdd{anton.chudaykin@unige.ch}
\emailAdd{ivanov99@mit.edu}
\emailAdd{kallosh@stanford.edu}
\emailAdd{alinde@stanford.edu}
\emailAdd{ophilcox@stanford.edu}
\emailAdd{yamada@ibs.re.kr}

\abstract{
We study the impact of spatial curvature ($\Omega_k$) and dynamical dark energy (parametrized by $w_0$ and $w_a$) on the spectral index $n_s$ using a combination of cosmic microwave background datasets (Planck, SPT, and ACT), and spectroscopic galaxy samples from DESI, including both BAO and full-shape clustering measurements. We show that a small negative curvature, $\Omega_k\simeq 3\times 10^{-3}$, lowers the value of $n_s$, bringing it  closer to predictions of the Starobinsky, Higgs, and simplest $\alpha$-attractor inflationary models. In particular, we find $n_s= 0.9667\pm0.0041$ (using Planck and DESI data) or $n_s= 0.9692\pm0.0035$ (adding ACT and SPT). Allowing for time-evolving dark energy also reduces the spectral index, leading to $n_s=0.9716\pm0.0032$ (from the combined dataset), or $n_s=0.9694\pm0.0035$ in combination with a small negative curvature.  Our results demonstrate that the tension between current observational data and the Starobinsky, Higgs, and simplest $\alpha$-attractor models holds only for $\Lambda$CDM, and can be mitigated in extended cosmological models. We discuss implications of these findings for inflationary models in an open universe and/or with dynamical dark energy, including scenarios with quantum tunneling and non-standard topology. Furthermore, we briefly describe a special class of $\alpha$-attractor models, where one can make $n_s$ arbitrarily large, and we describe the $\alpha$-attractor  quintessence model. Such models may be of particular relevance when future data from DESI, as well as DESI-II, SPHEREx, Euclid, Rubin, and Roman, becomes available.}

\preprint{MIT-CTP/6076}

\begin{document}

\maketitle

 
\parskip 7pt

\section{Introduction}

More than a decade ago, WMAP and Planck data releases identified two inflationary models that provided the best fit to their results: the Starobinsky model \cite{Starobinsky:1980te} and Higgs inflation \cite{Salopek:1988qh,Bezrukov:2007ep}. While very different in origin, the models yield the same predictions for the spectral index $n_{s}$ and the tensor-to-scalar ratio $r$ as a function of the number of e-foldings $N$. 

A further investigation, stimulated by the first Planck data release, led to the development of $\alpha$-attractors \cite{Kallosh:2013yoa}, which generalize the Starobinsky model and Higgs inflation.  The simplest $\alpha$-attractors, E-models and T-models, have potentials
\be\label{ET}
V_E= V_{0} (1-e^{-\sqrt{2\over 3\alpha}\vp})^{2n}\ , \qquad V_{T}= V_{0} \tanh^{2n}{\vp\over \sqrt{6\alpha}} \ .
\ee
For $n = 1$ and $\alpha = 1$, the potential $V_{E}$ coincides with the potential of the Starobinsky model, while for $n = 2$ and $\alpha = 1$, the potential $V_{T}$ is very similar to that of Higgs inflation. In all cases, at $\vp \gg \sqrt{3\alpha/2}$, the potentials have the same universal behavior 
\be
V= V_{0}(1- c\,e^{-\sqrt{2\over 3\alpha}\vp}+...)
\ee
where the constant $c$ can be absorbed in a redefinition (shift) of the field $\vp$. We call these models ``exponential attractors'' because their potentials approach the plateau exponentially fast, and their predictions in the small $\alpha$, large $N$ limit, do not depend on $n$: 
\be\label{apred}
n_{s} = 1-{2\over N}\ , \qquad r  = {12\alpha \over N^{2}} \ .
\ee
These results satisfy the universal consistency condition
\be\label{nr}
r = 3\alpha (1-n_{s})^{2} \ .
\ee
For $\alpha = 1$, these predictions coincide with the prediction for $n_{s}$ in the Starobinsky model and Higgs inflation; however, $\alpha$-attractors provide the possibility to fit the data with any value of the tensor-to-scalar ratio $r$ by a proper choice of $\alpha$. As a result, the Starobinsky model, Higgs inflation, and $\alpha$-attractors became the standard benchmarks for planning B-mode searches~\cite{Chang:2022tzj,LiteBIRD:2022cnt}. 

Notably, the required number of inflationary e-folds $N$ depends on reheating and the post-inflationary evolution. WMAP and Planck considered the range $ 50 \leq N \leq 60$. If, for example, one takes $N = 60$, one finds $n_{s} = 0.9667$, which is quite consistent with WMAP and Planck data. Lately, because of the new constraints on $r$ and the considerations related to reheating, expectations have gradually shifted towards the range $47 \leq N \leq 57$. Combining  CMB data from Planck and BICEP/Keck with the latest ACT and SPT data \cite{AtacamaCosmologyTelescope:2025blo,SPT-3G:2025bzu} increases $n_{s}$ to $n_{s}=0.9682\pm 0.0032$, which is still compatible with the benchmark models. 
The situation changes when one takes into account the recent DESI DR2 data \cite{DESI:2025zgx,AtacamaCosmologyTelescope:2025blo,SPT-3G:2025bzu,Balkenhol:2025wms}: the spectral index obtained in joint CMB + DESI fits is given by: $n_{s}=0.9728\pm0.0029$  \cite{SPT-3G:2025bzu,Balkenhol:2025wms}, which may rule out some of the benchmark models at the $\gtrsim 2\sigma$ level.

Here we should add an important clarification: one can easily increase the value of $n_{s}$ and match the CMB + DESI data in {\it two-field} $\alpha$-attractor models, such as hybrid $\alpha$-attractors \cite{Kallosh:2022ggf,Braglia:2022phb}. Thus the real question is the following: if the benchmark {\it single-field} models are disfavored when combining CMB data with DESI data, can one find other equally compelling {\it single-field} inflationary models? 

This issue has been explored in many papers following the recent ACT and SPT data releases \cite{AtacamaCosmologyTelescope:2025blo,SPT-3G:2025bzu}. While some of the most interesting suggestions are described in the review article \cite{Kallosh:2025ijd}, there have been many other, more recent efforts. One of the compelling suggestions concerned polynomial attractors (P-models), in which the potential approaches the plateau not exponentially but polynomially. 
The simplest examples include the potentials
\be
V =V_{0} {\phi^{k}\over \phi^{k} + \mu^{k}} =  V_{0}\left(1- { \mu^{k}\over \phi^{k}} +...\right).
\ee
Such potentials appear in a variety of different contexts, such as mutated hybrid inflation \cite{Stewart:1994pt}, brane inflation \cite{Dvali:1998pa}, KKLT-related brane inflation \cite{Kachru:2003sx,Martin:2013tda,Kallosh:2018zsi}, pole inflation \cite{Galante:2014ifa,Terada:2016nqg,Kallosh:2019hzo}, and also in polynomial $\alpha$-attractors and $\xi$-attractors \cite{Kallosh:2022feu,Kallosh:2026qrc,Kallosh:2026tke}.  These models predict that at small $\mu$ 
\be\label{KKLTIw}\
  n_{s} = 1-{2(k+1)\over (k+2)N} \, , 
\ee  
which spans the broad range of $n_{s}$ from $1-{2\over N}$ to $1-{1\over N}$. In particular, for $N = 55$, this gives the range $0.9636 < n_{s} < 0.9818$, which is more than sufficient to cover the CMB-DESI range $n_s= 0.9728\pm 0.0029$. 

Clearly, we have a variety of interesting models matching the CMB-DESI data. That said, it is still too early to give up on the favorites of the last decade, especially given that previous works \cite{SPT-3G:2025bzu,Ferreira:2025lrd,McDonough:2025lzo,Balkenhol:2025wms} have suggested a $\approx 3\sigma$ tension between DESI DR2 and CMB data in the $\Lambda$CDM model.
In particular, the new constraints  $n_{s}=0.9728\pm0.0029$ have been obtained in the context of $\Lambda $CDM in a flat universe, but this very context, according to DESI  DR2 \cite{DESI:2025zgx,Chudaykin:2025lww,Ivanov:2026dvl}, is disfavored at around $3\sigma$.\footnote{Some potential methodological concerns with respect to the DESI preference for dynamical dark energy over $\Lambda$CDM can be found in \cite{Ong:2025utx,Afroz:2025iwo,Ong:2026tta,Garcia-Garcia:2026nzy}.}

In this paper, we will examine observational constraints on $n_{s}$ while relaxing some of the standard  $\Lambda$CDM assumptions, and consider their theoretical context within inflationary model building. 
First of all, we will relax the condition of a flat Universe and consider what is known about the cosmological data with $\Omega_k\neq 0$ and how the non-vanishing 
$\Omega_k$ affects the value of $n_s$. Our interest is motivated by the study in \cite{Chen:2025mlf} where the role of the curvature in light of BAO
from DESI DR2 was investigated. In the former paper, the important role of curvature was explained; building on this, \cite{Chudaykin:2025lww} performed a detailed investigation folding in both BAO and full-shape spectroscopic data,\footnote{See \citep{Chen:2025mlf,Yadav:2025wbc,Giare:2026oti,Pulido-Hernandez:2026hcs,Wang:2025hvo,daCosta:2024grm,Wu:2024faw,DESI:2025zgx} for other discussions of curvature in light of DESI BAO data.} but the explicit effect of curvature on $n_s$ was not included.

Our observational results are discussed in \S\ref{sec: results} and summarized in Fig.~\ref{fig:SPAns}. Allowing for spatial curvature lowers the inferred value of $n_s$, bringing the simplest inflationary models with plateau potentials into better agreement with the observational data. In particular, in the analysis of Planck, ACT DR6 and SPT-3G primary CMB measurements, Planck+ACT CMB lensing, DESI DR2 BAO and DESI DR1 full-shape clustering statistics (as detailed in \S\ref{sec: data}), the value of $n_s$ is consistent with the predictions of Higgs and Starobinsky inflation within $1.5\sigma$ and $2.3\sigma$, respectively. Importantly, the current data favor a slightly open Universe at the $2.7\sigma$ level. In \S\ref{sec: theory} of this paper, we place these results in context by reviewing the cosmological scenario with non-zero spatial curvature and discuss the issue of consistent inflationary models in a slightly open Universe.  

We secondly explore the time-evolving dark energy scenario, which also predicts lower values of $n_s$, although the effect is more modest. In \S\ref{sec: theoryDE}, we describe new inflationary models, with or without dynamical dark energy, that feature flexible values of $n_s$ that can be used in the future when more experiments might converge to a particular value of $n_s$, including a new forecast for LiteBIRD in Fig.~\ref{Flux2}.
Finally, we discuss implications of the main results of this paper in \S\ref{sec: discussion}. 

\section{Datasets}\label{sec: data}
This work presents cosmological results in models featuring one or both of free spatial curvature and dynamical dark energy.
To obtain these, we follow the procedure of~\cite{Chudaykin:2025lww}, varying six cosmological parameters of the base $\Lambda$CDM model
($H_0,\omega_c\equiv \Omega_ch^2, \omega_b \equiv \Omega_bh^2, A_s, n_s,\tau$), as well as those describing beyond-the-standard-model extensions. These include a spatial curvature parameterized by 
its current contribution 
to the Friedmann equation.
$\Omk$ (with $\Omk>0$ indicating a negative curvature), and dynamical dark energy with the linear equation-of-state $w(a)=w_0+w_a(1-a)$.

To model the full-shape galaxy clustering statistics (introduced below), we utilize a theoretical model based on 
Large-Scale Structure 
Effective Field Theory as described in~\cite{Chudaykin:2025aux,Chudaykin:2025lww} (see also \citep[][]{Philcox:2021kcw,Ivanov:2021kcd,Ivanov:2019pdj,Chudaykin:2022nru,Chudaykin:2024wlw,Ivanov:2026dvl,Chen:2021wdi,DAmico:2019fhj}). The galaxy power spectrum (bispectrum) is computed at one-loop order (tree-level) in cosmological perturbation theory, including all relevant effects such as galaxy bias, resummation of the long-wavelength displacements, corrections for short-wavelength physics, and stochasticity. This is implemented in the \textsc{class-pt} code \citep{Chudaykin:2020aoj}.

We analyze the following datasets:
\begin{itemize}

\item \textbf{CMB:} We employ the high-$\ell$ \texttt{plik} TT, TE, and EE spectra, together with the low-$\ell$ \textsc{SimAll} EE and low-$\ell$ \textsc{Commander} TT likelihoods from the official Planck 2018 release~\cite{Planck:2018nkj}. In addition, we include measurements of the lensing potential auto-spectrum $C_L^{\phi\phi}$ from Planck's \texttt{NPIPE} maps and the Atacama Cosmology Telescope (ACT) Data Release 6 (DR6) \cite{Carron:2022eyg,ACT:2023kun,ACT:2023dou}.

\item \textbf{CMB\text{-}SPA:} 
We combine ground-based measurements of the TT/TE/EE (\textit{i.e.}\ T\&E) power spectra from ACT DR6~\cite{AtacamaCosmologyTelescope:2025blo} and SPT-3G D1~\cite{SPT-3G:2025bzu} with high-$\ell$ Planck data at $\ell<1000$ in TT and $\ell<600$ in TE/EE from the \texttt{plik\_lite} PR3 likelihood~\cite{Planck:2019nip}, together with low-$\ell$ \textsc{Commander} TT measurements~\cite{Planck:2019nip} and a Gaussian prior on the optical depth, $\tau=0.051\pm0.006$~\cite{Planck:2020olo} (encapsulating low-$\ell$ polarization data).
We use the compressed \texttt{ACT-lite} likelihood, which provides CMB-only bandpower estimates marginalized over foreground and most systematic parameters~\cite{AtacamaCosmologyTelescope:2025blo}. We combine these data with lensing potential measurements from Planck PR4 and ACT DR6, as in the previous case.

\item \textbf{BAO:} We include BAO measurements from DESI data release two (DR2) \citep{DESI:2025zpo,DESI:2025zgx}, based on post-reconstruction clustering measurements. These include distance measurements from the Bright Galaxy Sample (BGS) in the redshift range $0.1 < z < 0.4$, two Luminous Red Galaxy (LRG) samples in $0.4 < z < 0.6$ (LRG1) and
$0.6 < z < 0.8$ (LRG2), an Emission Line Galaxy (ELG) sample in $1.1 < z < 1.6$ (ELG2), a combined LRG and ELG tracer in $0.8 < z < 1.1$ (LRG3 + ELG1), and a quasar sample covering $0.8 < z < 2.1$ (QSO). We additionally include the BAO signal from the Lyman-$\alpha$ forest and its cross-correlation with quasars.

\item \textbf{Full-Shape Statistics ($P_\ell+B_0$):} We utilize the DESI data release one (DR1) redshift-space power spectrum multipoles ($\ell=0,2,4$) and bispectrum monopole measured from the public DESI catalogs~\cite{DESI:2025fxa,Chudaykin:2025aux}. These data comprise six non-overlapping spectroscopic data chunks: BGS, LRG1, LRG2, LRG3, ELG2, and QSO samples. As discussed in~\cite{desi1}, the two- and three-point statistics are measured using quasi-optimal estimators, using the \polybin implementation \citep{Philcox:2024rqr}, which accounts for various effects including fiber collisions and integral constraints. Data cuts and priors on the EFT parameters are described in~\cite{desi2}. We omit the cross-correlation between DR1 full-shape and DR2 BAO, as discussed in \citep{Chudaykin:2025aux} (see Appendix C), and validated in \citep{Forero-Sanchez:2026bff}.
The independent DESI likelihood has also recently been
extended to small scales with a consistent 
one-loop bispectrum model~\cite{Ivanov:2026dvl} (see also \citep{Bakx:2025pop,Philcox:2022frc,DAmico:2022ukl}). Given that its impact on the extended cosmological models considered in this work is quite limited, we use the tree-level bispectrum likelihood as a baseline in this study.

\item \textbf{SN:} We utilize Type Ia supernova data from the Pantheon+ sample~\citep{Brout:2022vxf}, comprising 1550
spectroscopically confirmed SNe in the redshift range
$0.001 < z < 2.26$. We adopt the public likelihood
from~\cite{Chudaykin:2025gdn} and marginalize over the supernova absolute magnitude analytically. 

\end{itemize}

\section{Results}\label{sec: results}

\begin{figure}[!t]
\begin{center}
\hskip -0.3cm
\includegraphics[width=0.45\textwidth]{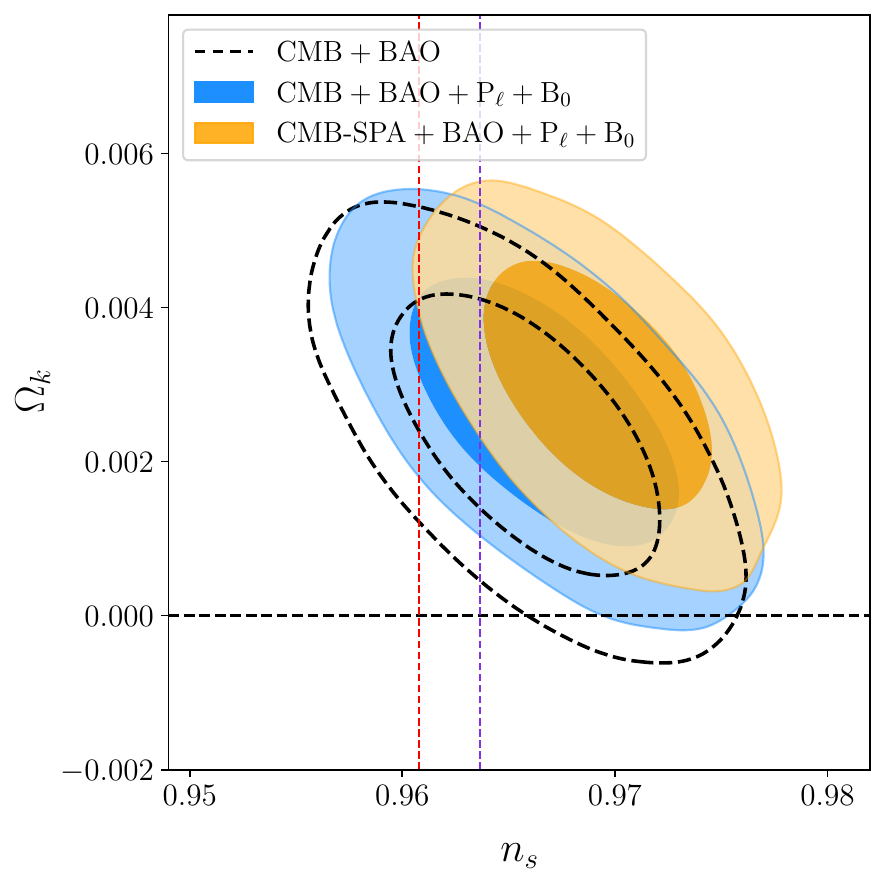}
\includegraphics[width=0.45\textwidth]{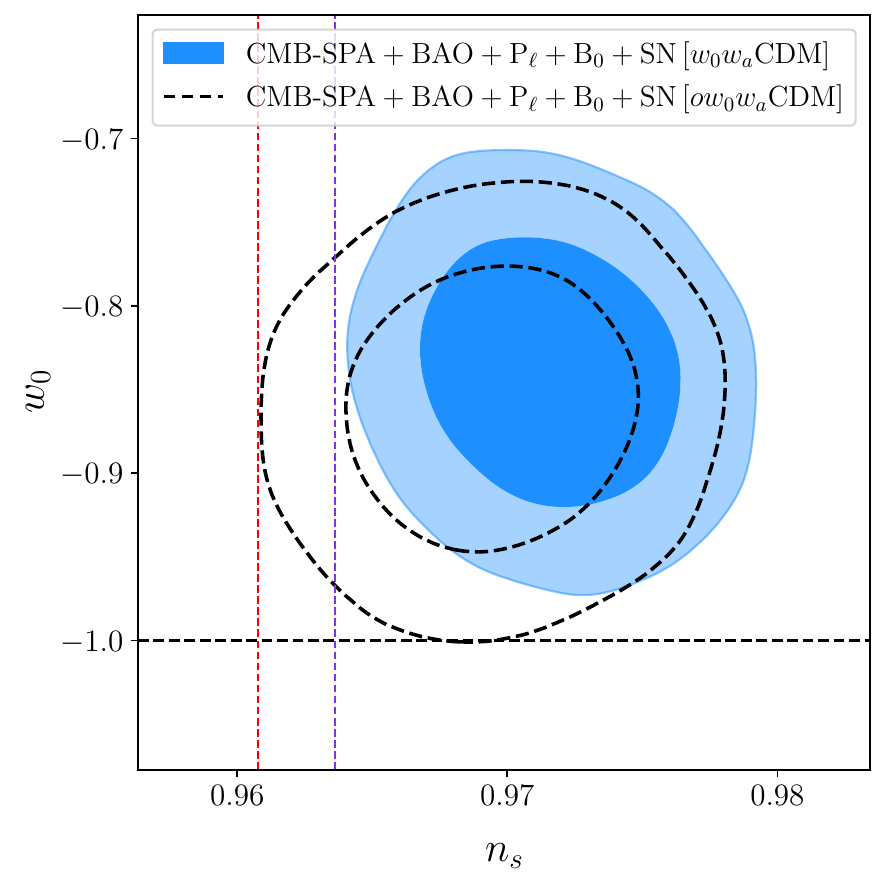}
\caption{\footnotesize {\it Left panel: } Joint constraints on $\Omk$ and $n_s$ from cosmological datasets in the $\OmK$ model. The data favor an open Universe with $\Omega_k \simeq 3\times 10^{-3}$, leading to $n_s\approx0.969$, which is $1.8\sigma$ lower than the corresponding $\ld$ result and consistent with the theoretically motivated single-field inflationary models. 
The $\cmbspa+\bao+P_\ell+B_0$ analysis (yellow) indicates a preference for $\Omk>0$ at the level of $2.7\sigma$.
{\it Right panel}: Joint constraints on $w_0$ and $n_s$ derived from the $\cmbspa+\bao+P_\ell+B_0+\sn$ dataset in the $\wa$ and $\owa$ scenarios. While the value of $w_0$ is fairly insensitive to curvature, we find that free $\Omk$ allows for lower values of $n_s$, in closer agreement with inflationary models. The dashed red and purple vertical lines mark the theoretical predictions for the Starobinsky and Higgs inflation models, respectively.} 
\vskip -15pt
\label{fig:SPAns}
\end{center}
\end{figure}

Next, we discuss our results and their implications for curvature and dark energy. 
Our constraints on cosmological parameters are summarized in Tab.~\ref{tab:param}, with two-dimensional posterior distributions in the $\Omk-n_s$ and $w_0-n_s$ planes shown in Fig.~\ref{fig:SPAns}. 
Tab.~\ref{tab:chi2} summarizes the significance of the preference for extended models over $\ld$, using different statistical metrics. 
In the literature there is a vast landscape of inflationary models; in the text, we choose to compare to a set of reference models, namely Starobinsky $R^2$ inflation with $N_*=51$~\cite{Starobinsky:1980te,Mukhanov:1981xt,Starobinsky:1983zz} and Higgs inflation with $N_*=55$~\cite{Bezrukov:2007ep,Bezrukov:2011gp}. Further discussion of the implications of our results on inflationary models can be found in \S\ref{sec: implications}.

\begin{table*}[!t]
    \centering
    \footnotesize
    \setlength{\tabcolsep}{3pt}
    \begin{tabular}{lccccccc}
    \toprule
    Dataset 
    & $10^3\Omk$ 
    & $n_s$ 
    & $\Omega_m$ 
    & $H_0$ 
    & $w_0$
    & $w_a$ 
    \\
    \midrule
    $\bm{\Lambda}$\textbf{CDM} &  &  &  &  &  & & \\
    $\cmb+\bao$ 
    & $-$ 
& $0.9707_{-0.0033}^{+0.0033}$ 
& $0.3012_{-0.0037}^{+0.0037}$ 
& $68.39_{-0.29}^{+0.29}$ 
& $-$
& $-$
\\
    $\cmb+\bao+P_\ell+B_0$ 
    & $-$ 
& $0.9723_{-0.0033}^{+0.0033}$ 
& $0.2984_{-0.0034}^{+0.0034}$ 
& $68.60_{-0.27}^{+0.27}$ 
& $-$
& $-$
\\
    $\cmbspa$ 
    & $-$ 
& $0.9687_{-0.0033}^{+0.0034}$ 
& $0.3161_{-0.0058}^{+0.0058}$ 
& $67.28_{-0.40}^{+0.40}$ 
& $-$
& $-$
\\
    \textcolor{blue}{$\cmbspa+\bao+P_\ell+B_0$} 
    & $-$ 
& \textcolor{blue}{$0.9745_{-0.0029}^{+0.0029}$} 
& \textcolor{blue}{$0.3005_{-0.0032}^{+0.0032}$} 
& \textcolor{blue}{$68.38_{-0.24}^{+0.24}$} 
& $-$
& $-$
\\
    \midrule
    $\bm{\OmK}$ &  &  &  &  &  &  & \\
    $\cmb+\bao$ 
    & $2.4_{-1.2}^{+1.2}$ 
& $0.9659_{-0.0042}^{+0.0042}$
& $0.3031_{-0.0038}^{+0.0039}$ 
& $68.64_{-0.32}^{+0.32}$ 
& $-$
& $-$
    \\
    $\cmb+\bao+P_\ell+B_0$ 
    & $2.7_{-1.2}^{+1.2}$ 
& $0.9667_{-0.0042}^{+0.0041}$
& $0.3007_{-0.0036}^{+0.0036}$ 
& $68.87_{-0.31}^{+0.31}$ 
& $-$
& $-$
    \\
    \textcolor{blue}{$\cmbspa+\bao+P_\ell+B_0$}
& \textcolor{blue}{$3.0_{-1.1}^{+1.1}$} 
& \textcolor{blue}{$0.9692_{-0.0035}^{+0.0036}$} 
& \textcolor{blue}{$0.3013_{-0.0033}^{+0.0033}$} 
& \textcolor{blue}{$68.86_{-0.31}^{+0.31}$} 
& $-$
& $-$
    \\
    \midrule
    $\bm{w_0w_a}$\textbf{CDM} &  &  &  &  &  &  & \\
    \textcolor{blue}{$\cmbspa\!+\!\bao\!+\!P_\ell\!+\!B_0\!+\!\sn$} 
& $-$ 
& \textcolor{blue}{$0.9716_{-0.0032}^{+0.0032}$} 
& \textcolor{blue}{$0.3085_{-0.0054}^{+0.0054}$} 
& \textcolor{blue}{$67.80_{-0.56}^{+0.56}$} 
& \textcolor{blue}{$-0.839_{-0.054}^{+0.054}$} 
& \textcolor{blue}{$-0.61_{-0.19}^{+0.19}$} 
    \\
    \midrule
    $\bm{ow_0w_a}\bm{{\rm CDM}}$ &  &  &  &  &  &  & \\
    \textcolor{blue}{$\cmbspa\!+\!\bao\!+\!P_\ell\!+\!B_0\!+\!\sn$} 
& \textcolor{blue}{$1.7_{-1.3}^{+1.3}$} 
& \textcolor{blue}{$0.9694_{-0.0035}^{+0.0035}$} 
& \textcolor{blue}{$0.3089_{-0.0054}^{+0.0053}$} 
& \textcolor{blue}{$68.00_{-0.57}^{+0.59}$} 
& \textcolor{blue}{$-0.861_{-0.056}^{+0.055}$} 
& \textcolor{blue}{$-0.50_{-0.21}^{+0.21}$} 
    \\
    \bottomrule
    \end{tabular}
    \caption{
    \footnotesize 
    Constraints on cosmological parameters in the standard $\ld$ model (first panel), a model with free spatial curvature ($\OmK$; second panel), one with dynamical dark energy ($\wa$; third panel), and the combined extension ($\owa$; last panel). We quote the mean and 68\% confidence intervals for all parameters. Note that supernovae data are included only when analyzing dynamical dark energy models (where low-redshift calibration is desirable).
    The rows shown in blue represent the baseline analyses of this work. 
    Two-dimensional marginalized constraints on $\Omk$, $n_s$, $w_0$ are shown in Fig.~\ref{fig:SPAns}. 
    }
    \label{tab:param}
\end{table*}

\subsection{Baseline}
First, we consider the $\ld$ model. We find that including the DESI DR1 full-shape data in the analysis leads to a modest increase of $0.5\sigma$ in the spectral tilt $n_s$ relative to the CMB-plus-BAO results (used in \citep{DESI:2025zgx} and a variety of other works), see Tab.~\ref{tab:param}. 
The broadband shape of the galaxy power spectrum and bispectrum affects the inference of 
the spectral tilt of the primordial power spectrum
in combination with the CMB 
mostly by exploiting the 
CMB 
$\omega_{\rm cdm}-n_s$ degeneracy.
Lower values of $\omega_{\rm cdm}$
preferred by the DR1 full-shape data~\cite{Chudaykin:2025aux} pull $n_s$ upwards
along the negative degeneracy direction. 
When the ground-based CMB measurements from ACT DR6 and SPT-3G D1 are further included, the mean value of $n_s$ increases by an additional $0.5\sigma$, yielding $n_s=0.9745\pm0.0029$. 
Our $\ld$ results are in good agreement with the recent Planck+ACT+SPT+BAO analysis~\cite{Balkenhol:2025wms}, which reported $n_s = 0.9728 \pm 0.0029$; however, we obtain a slightly higher value of $n_s$ owing to the inclusion of the DESI DR1 full-shape data. 
The combined result is in nominal tension with the predictions of Starobinsky and Higgs inflation at the $4.7\sigma$ and $3.7\sigma$ levels, respectively, making these scenarios statistically disfavored.

\begin{table}
\centering
\small
    \begin{tabular}{lccc}
    \toprule
    Models 
    & $\Delta\chi^2_{\rm MAP}$ 
    & Significance 
    & $\rm \Delta (AIC)$ 
    \\
    \midrule
    $\OmK$ 
    & $-7.3$ 
    & $2.7\sigma$
    & $-5.3$ \\
    $\wa$ 
    & $-13.3$ 
    & $3.2\sigma$ 
    & $-9.3$ \\
    $\owa$ 
    & $-15.0$ 
    & $3.1\sigma$ 
    & $-9.0$ \\
    \bottomrule
    \end{tabular}
\caption{\footnotesize Preference for the alternative cosmological models compared to $\ld$. The second column reports the difference in the $\chi^2_{\rm MAP}$ for the best-fit model relative to the best-fit $\ld$ model. The third column lists the corresponding (frequentist) significance levels given extra free parameters, and the final column gives the difference in the Akaike Information Criterion value~\cite{AIC}, $\rm \Delta (AIC)=AIC_{\rm model}-AIC_{\ld}$, where ${\rm AIC} = \chi^2_{\rm MAP}+ 2N_p$ and $N_p$ is the number of free parameters in the model. According to the commonly adopted (but heuristic) interpretation of \cite{Burnham2002}, $-2<\Delta ({\rm AIC})<0$ indicates weak support for the extended scenario, $-6<\Delta ({\rm AIC})<-2$ indicates moderate support, and $\Delta ({\rm AIC})<-6$ indicates strong support relative to $\ld$.
The results are reported for the baseline analyses of this work (highlighted in blue in Tab.~\ref{tab:param}).
\label{tab:chi2}
}
\end{table}

\subsection{Adding Curvature}
Next, we explore the $\OmK$ scenario, featuring free spatial curvature (see the second panel of Tab.~\ref{tab:param}). 
Consistent with previous studies~\cite{DESI:2025zgx,desi2}, we find that the joint analysis of CMB and DESI data prefers a slightly open universe: the $\cmb+\bao$ data favor $\Omk>0$ at the $1.9\sigma$ level, while the inclusion of the DESI DR1 full-shape measurements increases the preference to $2.4\sigma$.\footnote{In what follows, we compute the significance of parameter deviations using the full posterior distribution, without assuming Gaussianity.} Including the ground-based CMB measurements from ACT DR6 and SPT-3G D1 strengthens this preference further to $2.7\sigma$. In the combined $\cmbspa+\bao+P_\ell+B_0$ analysis, we find
\be
\Omega_k= (3.0\pm 1.1)\times 10^{-3}\,
\ee
at 68\% confidence level.
As shown in Fig.~\ref{fig:SPAns}, $\Omk$ and $n_s$ are anticorrelated, so a more negative curvature ($\Omk>0$) leads to lower values of $n_s$. Consequently, the results become more consistent with the theoretically motivated inflationary scenarios. In the $\cmbspa+\bao+P_\ell+B_0$ analysis, we obtain
\be
n_s=0.9692\pm0.0035\,,
\ee
which is $1.8\sigma$ lower than the value obtained in the $\ld$ model. This result is consistent with Higgs and Starobinsky inflation within $1.5\sigma$ and $2.3\sigma$, respectively. We conclude that a slightly negative curvature $\Omega_k \simeq 3\times 10^{-3}$ can reconcile the simplest theoretically motivated inflationary models with the observational data, restoring their viability.

Notably, the $\OmK$ extension reduces the best-fit $\chi^2$ of the model by $7.3$ units compared to $\ld$ (see Tab.~\ref{tab:chi2}), which corresponds to $2.7\sigma$. The associated $\Delta ({\rm AIC})=-5.3$ indicates that the open Universe scenario provides a substantially better description of the data after accounting for its additional free parameter.

\subsection{Adding Dynamical Dark Energy}
Next, we consider the $\wa$ model, including dynamical dark energy. Similar to curvature, evolving dark energy reduces the value of the spectral tilt $n_s$, making the results more consistent with the physically motivated inflationary models than those in $\ld$. In particular, the $\cmbspa+\bao+P_\ell+B_0+\sn$ analysis\footnote{We include the SNe measurements to further constrain the expansion history in the $\wa$ background~\cite{DESI:2024hhd,desi2}.} yields $n_s=0.9716\pm0.0032$, which is $1.0\sigma$ lower than the corresponding value in the $\ld$ model.
This matches expectations, given that (a) the cosmological data favor a quintessence-like dark energy equation-of-state parameter at the present epoch~\cite{DESI:2025zgx,DESI:2025fii}, \textit{i.e.}\ $w_0>-1$, and (b) $w_0$ is anticorrelated with $n_s$ (see Fig.\,\ref{fig:SPAns}). Notably, the corresponding shift in $n_s$ is smaller than that found in the free curvature scenario due to the weaker correlation between $w_0$ and $n_s$ in the dynamical dark energy model (see the right panel of Fig.~\ref{fig:SPAns}). In particular, the $\wa$ constraint on $n_s$ is in $3.4\sigma$ and $2.5\sigma$ tension with the prediction of Starobinsky and Higgs inflation, respectively. In conclusion, time-evolving dark energy cannot fully reconcile Starobinsky inflation with the observational data, although Higgs inflation remains viable. As for $\OmK$, we find a clear preference of the data for $\wa$ over $\ld$, with an improvement of $13.3$ in the best-fit $\chi^2$ (corresponding to $3.2\sigma$) and a $\Delta(\rm AIC)=-9.3$.

\subsection{Curvature \& Dynamical Dark Energy}
Finally, we explore the combined scenario: $\owa$, featuring both free curvature and time-evolving dark energy. We find that the $\cmbspa+\bao+P_\ell+B_0+\sn$ data yield $n_s=0.9694\pm0.0035$, which is very similar to the result obtained in the model with free $\Omk$ alone. Allowing for free curvature alters the degeneracy direction in the $w_0-n_s$ plane (see Fig.~\ref{fig:SPAns}), resulting in a weak positive correlation between the two parameters. This shifts $n_s$ toward smaller values, improving the agreement with the Starobinsky and Higgs inflationary models.
It is worth noting that the tension between the CMB and DESI BAO data is already alleviated by introducing dynamical dark energy~\cite{DESI:2025zgx} (though this extension is certainly non-trivial from a theoretical perspective). Consequently, the preference for an open universe in a dynamical dark energy background is reduced from $2.7\sigma$ to $1.5\sigma$; this reduces the statistical evidence for non-zero spatial curvature.

The conclusion is clear: allowing for free spatial curvature is sufficient to accommodate lower values of $n_s$ consistent with Starobinsky and Higgs inflation, while the addition of dynamical dark energy has only a marginal impact on the inferred $n_s$ value. The data show a similar preference for the $\owa$ scenario as for $\wa$, with $\chi^2$ improving by $15.0$ units, corresponding to $3.1\sigma$ levels, with $\Delta ({\rm AIC})=-9.0$. We do not find a strong preference for $\owa$ over $\wa$.

Future data is expected to significantly sharpen the constraints on both curvature and dynamical dark energy. As shown above, the DESI full-shape clustering data significantly impacts the $n_s$ posterior, and we expect yet greater improvements with the release of DR2 clustering, alongside future BAO datasets. Moreover, the Euclid photometric and spectroscopic datasets, expected in late 2027, will provide crucial independent handles on the late Universe evolution, through lensing, BAO, and clustering probes.

\section{Open Universe and Inflation: A Short Overview}\label{sec: theory}
 \subsection{Implications for Models of Inflation}\label{sec: implications}

Below, we comment on the feasibility of inflationary models, given the $n_s$ constraints presented in \S\ref{sec: results}. Here, it is useful to distinguish between Starobinsky inflation, Higgs inflation, and the broader class of exponential $\alpha$-attractors.  These models share the same form of the spectral index at leading order:
\begin{align}
    n_s \simeq 1-\frac{2}{N};
\end{align}
as such, the main differences among these models arise from the allowed values of $N$.

As shown in Tab.~\ref{tab:param}, the combined CMB and DESI full-shape data prefer a relatively high value of the scalar spectral index $n_s\approx 0.9745$ in the flat $\ld$ interpretation.  This creates tension with the simplest exponential plateau models, such as Starobinsky inflation, Higgs inflation, and exponential $\alpha$-attractors, unless an extremely large value of $N\simeq 75-80$ is assumed. However, such a large $N$ cannot be realized in most inflationary models.

When spatial curvature is allowed, however, the inferred value of $n_s$
is shifted to smaller values, and from Tab.~\ref{tab:param}, we find 
$$
    n_s = 0.9692 \pm 0.0035 ,
$$
which substantially reduces the apparent tension with plateau models. Nevertheless, since the value of $N$ is not arbitrary, especially for Starobinsky inflation and Higgs inflation, it is useful to comment on the compatibility of each class of models separately.

For Starobinsky inflation, the post-inflationary dynamics is relatively rigid.  The shape of the potential after inflation is essentially fixed, and the scalaron oscillates around the minimum before reheating.  The intermediate epoch is therefore approximately matter dominated, and it is difficult to invoke a prolonged kination or stiff-reheating phase
which would significantly increase $N$.  Thus the reference value $N\simeq 51$ gives a
meaningful comparison~\cite{Gorbunov:2010bn,Bezrukov:2011gp}.  This reference value is not at the center of the open-universe posterior, but the tension is reduced to the level of about $2.3\sigma$.

Higgs inflation is also relatively constrained.  Since the inflaton is
identified with the Standard Model Higgs field, its couplings to Standard
Model particles are already present, and a long kination epoch is not
naturally expected~(see \citep[e.g.,][]{Garcia-Bellido:2008ycs,Bezrukov:2008ut} for reheating in Higgs inflation).  The conventional prediction, corresponding roughly
to  $N \simeq 55 ,$ lies within about $1.5\sigma$ of the open-universe result.
Thus Higgs inflation becomes reasonably compatible once spatial curvature is allowed.

By contrast, $\alpha$-attractors form a broader class of models.  The inflationary plateau is universal in the large-field regime, but the shape of the potential after inflation and the reheating sector are more model-dependent.  Therefore, a stiff or kination-like post-inflationary epoch may be possible in some realizations, allowing a somewhat larger
value of $N$ (see \citep[e.g.,][]{Iacconi:2025odq}). For a plateau-like prediction with $N=60$, one obtains
\begin{align}
    N = 60
    \qquad \Rightarrow \qquad
    n_s \simeq 0.9667 .
\end{align}
This is only about $0.7\sigma$ away from the open-universe value $n_s=0.9692\pm0.0035$.  Therefore, $\alpha$-attractors with moderately large $N$ can be fully compatible with the open-universe result.
Note also that $\alpha$-attractors include models beyond the exponential
plateau class.  In particular, polynomial $\alpha$-attractors~\cite{Kallosh:2022feu} can give
somewhat larger values of $n_s$ even without invoking an unusually large
$N$.

Based on the above, we do not interpret the results of \S\ref{sec: results} as selecting a specific
inflationary model.  Rather, they show that the apparent tension of
plateau models in the flat $\Lambda$CDM analysis is not robust against
extensions of the late-time cosmological model, especially spatial curvature. Starobinsky and Higgs inflation provide representative values because their post-inflationary histories are comparatively
restricted, whereas $\alpha$-attractors should be interpreted as a wider
model class with additional freedom in both the potential and reheating
history.

\subsection{Tunneling}
It is well known that $\Omega\equiv 1-\Omk \approx 1$ is one of the standard predictions of inflationary cosmology.   However, in the mid-1990s, prior to the discovery of dark energy, the observational data seemed to convincingly demonstrate that $\Omega \sim 0.3$. An attempt to address this problem was made in~\cite{Bucher:1994gb}. The authors suggested that an open universe could appear after the first stage of inflation as a result of the Coleman-De Luccia (CDL) tunneling with bubble formation \cite{Coleman:1980aw}. If inflation continues inside the bubble, one would be able to have $\Omega \sim 0.3$ inside the bubble universe for a certain choice of the potential.

However, the slow-roll condition  $|V''| \ll V$ implies that the potential must be very soft on both sides of the potential barrier. Meanwhile, the CDL tunneling usually requires that the slow-roll condition $|V''| \ll V$ has to be strongly violated across the barrier.\footnote{If the slow-roll condition  $|V''| \ll V$ is satisfied across the barrier, then tunneling occurs due to the accumulation of quantum fluctuations, gradually bringing the field to the top of the barrier  \cite{Linde:1990flp}, which triggers the process of eternal inflation in the observable part of the universe \cite{Vilenkin:1983xq,Linde:1986fd}. } In practical terms, this means that the slow-roll inflationary potential must have a sharp peak at a distance corresponding to $\sim 50$ e-foldings from the end of inflation, which has been difficult to realize in a physically motivated manner.
\begin{figure}[H]
\centering
		 \includegraphics[width=0.62\textwidth]{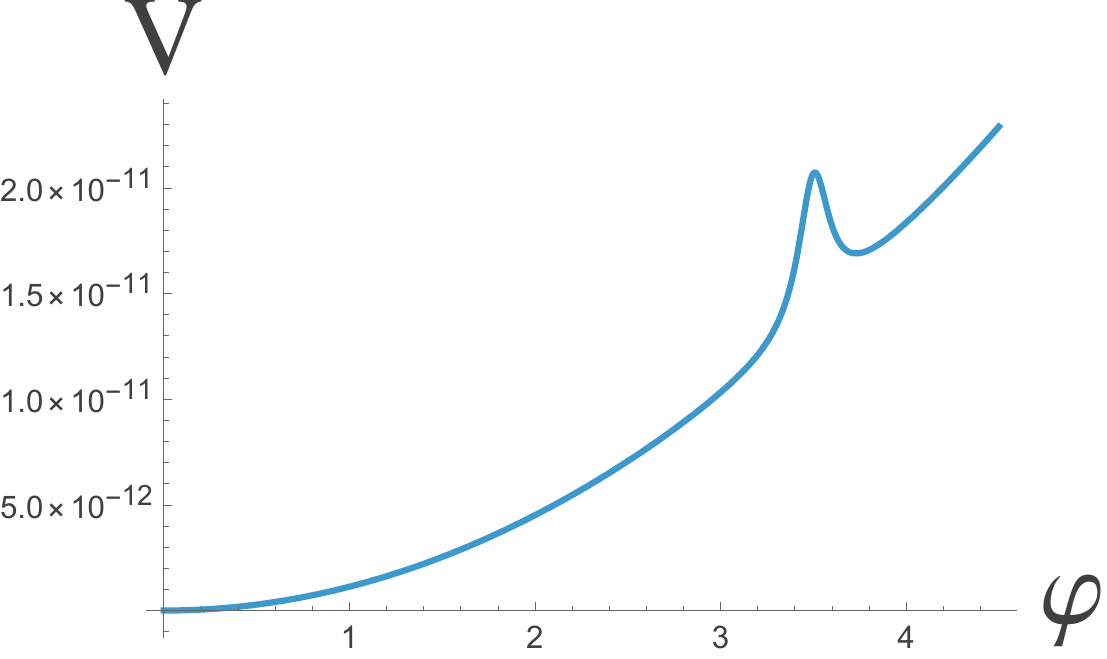}
        \caption{\footnotesize{The potential \ref{peakpot}  shown in Fig.1 in  \cite{Linde:1998iw}, but here the field $\phi$ is shown in reduced Planck mass units $M_{p}=1/\sqrt{8\pi G}$, and the potential is shown in units of $M^{4}_{p}$}  }  \label{Peak}    \end{figure}

The first model with the required properties was proposed in \cite{Linde:1998iw} and had the simplest chaotic inflation potential ${m^{2}\over 2}\phi^{2}$, plus a  peak located at $\phi = \phi_{0}$:
\be\label{peakpot}
V= {m^{2}\over 2}\phi^{2}\left(1+{a^{2}\over b^{2}+ (\phi-\phi_{0})^{2}}\right) 
\ee
with $a, b \ll 1$ and $\phi_{0} \sim 10$ in Planck mass units.
The resulting inflaton potential has a sharp peak and looks rather exotic, see Fig.~\ref{Peak}. However, the model works; see detailed investigations of scalar and tensor perturbations in this model and in several similar models in \cite{Sasaki:1998ug, Linde:1999wv, Yamauchi:2009zz, Yamauchi:2011qq,Bousso:2013uia}.  A common feature of the spectrum of inflationary perturbations in such models is a brief stage of fast rolling of the field immediately after the tunneling. This results in a suppression of the low-$\ell$ modes of the CMB spectrum, which could partially account for the observed suppression of these modes in the WMAP and Planck results \cite{Linde:1998iw} (see \cite{Contaldi:2003zv} for a subsequent discussion of a closely related effect in a flat universe).

After the discovery of dark energy, which brought $\Omega$ very close to 1, such models were largely forgotten, only to resurface when ideas from the string theory landscape drew attention to bubble formation and to open universes born due to tunneling between different vacua. An important observation in this respect was made by Freivogel,  Kleban,  Rodriguez Martinez, and Susskind \cite{Freivogel:2005vv}. In accordance with certain string-theory suggestions, the authors assumed that long stages of inflation are very rare in the landscape, implying that the number of e-foldings $N$ must be small. On the other hand, if $N$ is too small, the late stages of the evolution of the universe formed in the open universe bubble will be dominated by the combination of the positive cosmological constant and the slowly-decreasing term $1/a^{2}$ due to the spatial curvature. This would prevent galaxy formation, just as in the cosmological constant case studied in \cite{Linde:1984ir,Sakharov:1984csx,Weinberg:1987dv}.  Thus, the number of e-foldings cannot be too large or too small. Under some specific assumptions made in 
\cite{Freivogel:2005vv} the authors suggested that the universe is expected to be open, with
 \be
4\times 10^{-4} \, \lesssim \,  \Omega_k \, \lesssim \, 2 \times 10^{-2} \ .
\label{an} \ee
It is amazing to see in Fig.~\ref{fig:SPAns} that the current data support the case of $2.7\times 10^{-3}$, which is right in the middle of the anthropic prediction shown in \rf{an}.

As we argued, this scenario requires tunneling through a sharp barrier. There are no such barriers in the original versions of the Starobinsky model, Higgs inflation, or the simplest versions of $\alpha$-attractors \rf{ET}, though one can try to modify these models accordingly. In particular, it may be possible to add a sharp peak to $\alpha$-attractor potentials \rf{ET}, just as we did in \rf{peakpot}. Such models would not win a beauty contest, but they may succeed in the difficult task of creating an infinitely large, open inflationary universe. 

\subsection{Chaotic Inflation \& Topology}
Producing open universes through tunneling is not the only way to account for the possibility of a tiny, but non-zero, value of $\Omega_{k}$. For example,
one may consider the possibility of creating {\it an infinitely large open universe} within quantum cosmology \cite{Hawking:1998bn}. However, this mechanism is rather controversial \cite{Linde:1998gs, Bousso:1998ed, Vilenkin:1998pp, Vilenkin:1998rp}; so we will not discuss it here.
Another speculative but intriguing possibility is quantum creation of {\it a compact universe} with a non-trivial topology \cite{Zeldovich:1984vk,Coule:1999wg,Linde:2004nz}. These calculations suggest that the probability of quantum creation of flat and open universes is greater than the probability of quantum creation of closed universes \cite{Linde:1983mx,Linde:1984ir,Vilenkin:1984wp}. 

There are other related discussions of the importance of a small negative curvature, which may have implications for pre-inflationary dynamics. For example, in curvature-assisted compactification~\cite{Yamada:2026ipx}, negative curvature can act as a slowly redshifting component that helps trap moduli or radion fields before a subsequent four-dimensional inflationary phase dilutes the curvature remnant. This provides a complementary perspective that open curvature may have played a dynamical role before the observable stage of inflation.

In this paper, we take a simpler approach based on the chaotic inflation scenario \cite{Linde:1983gd}.
The basic idea of chaotic inflation is that if inflation begins in a sufficiently large part of the universe of size $\mathcal{O}(H^{-1})$ where the energy density is dominated by a sufficiently flat potential $V(\phi)$, it enters the inflationary regime, which quickly removes all previously existing inhomogeneities. In particular, if the potential is sufficiently flat at the Planck density $V=\mathcal{O}(1)$ in units $M_p^4$, 
then inflation can begin if the potential $V(\phi)$ dominates the energy density in a single domain of the smallest possible size $\mathcal{O}(1)$ (\textit{i.e.}\ the Planck length $M_p^{-1}$).
This result has been confirmed by numerical investigations \cite{Corman:2022alv}, as well as by arguments based on quantum cosmology \cite{Linde:1983mx,Linde:1984ir,Vilenkin:1984wp}.\footnote{Strictly speaking, the Planck density is $\mathcal{O}(1)$ if the Planck density is defined as $\rho_{P} = \mathcal{O}(G^{-2}) = \mathcal{O}(M_{p}^4)$. However, it is customary now to express values of the fields and their masses in terms of the reduced Planck mass $M_{p} = 1/\sqrt{8\pi G}$, and we do the same in this article. The definition of the Planck density should not depend on the choice of the system of units. As a useful guide, one can use the expression for the probability of quantum creation of an inflationary universe $P \sim \exp(-{3\over 8 G^{2} V})$ \cite{Linde:1983mx,Linde:1984ir,Vilenkin:1984wp}. This suggests that quantum creation of the universe is not suppressed for $V\gtrsim G^{-2}$, and quantum effects may become uncontrollably large at density greater than   $\rho_{P} = \mathcal{O}(G^{-2})$. That is why, long ago, a popular definition of the Planck mass was $M_{p} = 1/\sqrt{G}$, and the Planck density $\rho_{P} = \mathcal{O}(M_{p}^{4})$.  However, in terms of the reduced Planck mass $M_{p}= 1/\sqrt {8 \pi G}$, the Planck density becomes $\rho_{P} = (8\pi)^{2}\, \mathcal{O}(M_{p}^{4})$. Thus, if one uses the system of units $M_{p} = 1/\sqrt{8\pi G} = 1$, the Planck density in these units is not $\mathcal{O}(1)$, but rather  $\mathcal{O}((8\pi)^{2}) \gg 1$. This is just a result of the redefinition of $M_{p}$, which does not affect the final results.}

What if inflation is possible only for $V \sim 10^{{-10}}$ as in the Starobinsky model, Higgs inflation, and $\alpha$-attractors? One can always solve this problem by adding another scalar field that dominates at the first stage of inflation, beginning at $V=\mathcal{O}(1)$, and then vanishes, allowing the second stage of inflation driven by $\alpha$-attractors to begin \cite{Linde:2014nna,Carrasco:2015rva,Dimopoulos:2016yep,Linde:2017pwt}. However, it has recently been realized that in a compact universe, one can achieve the same result without assuming that the first stage of inflation begins at the Planck density.

Back in 1996, Cornish, Spergel, and Starkman \cite{Cornish:1996st} suggested considering models of a compact open or flat universe. If, for example, a flat universe is described by a torus, then one may begin with a flat universe of the Planck size. The Friedmann equations in such a space would be the same as usual, but the non-trivial topology may lead to interesting effects, such as chaotic mixing, that could ensure some degree of homogenization of the universe and its continuous expansion until the low-energy-scale inflation begins. If inflation lasts sufficiently long, it may erase all traces of the initial anisotropy of the universe, but if inflation is short, we may find ourselves in a slightly open and slightly anisotropic universe.

About 10 years later, a closely related scenario was investigated in \cite{East:2015ggf}, considering the evolution of a flat universe with a toroidal topology (a box with opposite sides identified). It was assumed that the universe was initially expanding, highly inhomogeneous, and dominated by gradient and kinetic energy. One might expect these inhomogeneities to prevent inflation. However, if the initial scalar field variations are contained within a sufficiently flat region of the inflaton potential, and the compact universe is spatially flat or open on average, then, according to \cite{East:2015ggf}, after the dilution of the gradient and kinetic energy due to expansion, the universe enters the inflationary regime. This happens even when overdense regions collapse to form black holes, because underdense regions continue to expand, allowing inflation to eventually begin.  A qualitative discussion of these numerical results can be found in \cite{Linde:2017pwt}.

These conclusions have been confirmed and generalized by numerous subsequent investigations \cite{Clough:2016ymm,Clough:2017efm,Aurrekoetxea:2019fhr,Creminelli:2020zvc,Joana:2020rxm,Joana:2022pzo,Corman:2022alv,Elley:2024alx,Aurrekoetxea:2024mdy,Joana:2024ltg}. Whereas these studies were mostly dedicated to flat compact universes, we expect that the results should be even stronger in an open universe case. Indeed, the energy density of matter and radiation in an expanding universe decreases much faster than the curvature contribution to the Einstein equation $\sim 1/a^{2}$, making the formation of collapsing parts of the universe even less probable. Furthermore, if the expanding universe does not collapse as a whole, then it continues to expand, the energy density and the curvature term $\sim 1/a^{2}$ decay, and the low-energy scale inflation begins, as in  \cite{East:2015ggf}.

One can implement this scenario in many inflationary models, including the Starobinsky model, Higgs inflation, and $\alpha$-attractors. But this, of course, brings us to the same issue as before: if the inflationary stage is very long, we will never see any remnants of the universe's initial openness and non-trivial topology. That said, perhaps arguments from string theory and the swampland conjectures disfavor long inflation. On the other hand, if inflation is too short, $\Omega_{k}$ is too large, and galaxies do not form \cite{Freivogel:2005vv}. This may explain why $\Omega_{k}$ is small but nonzero, which could make the large-scale anisotropy associated with the non-trivial topology of the universe small yet potentially observable.  

\section{Inflationary Models with   Flexible \boldmath{$n_s$} and Dark Energy }\label{sec: theoryDE}
\subsection{\boldmath{Waterfall-modulated   $\alpha$-attractor models}}
There are many ways to increase $n_s$ in inflationary models, in particular in $\alpha$-attractors, as we discussed in Introduction. Here we would like to introduce the waterfall-modulated   $\alpha$-attractor models, with flexible values of $n_s$, which are also useful for the models with dark energy.

We do not know which values of $n_s$ will be favored by future observations.  We therefore describe below a mechanism that allows one to change $n_s$ continuously, based on an idea of a premature end of inflation. This mechanism was originally introduced in two-field hybrid inflation \cite{Linde:1993cn,Kallosh:2022ggf}, where it was shown that the premature termination of inflation by the waterfall field can lead to a significant (and controllable) increase in $n_s$. Hybrid $\alpha$-attractors \cite{Kallosh:2022ggf,Braglia:2022phb}, like all other hybrid inflation models, have a waterfall stage driven by a second field, often called the waterfall field. The potential of the inflaton field is uplifted by the contribution of the waterfall field, and inflation ends not at the end of the slow-roll regime for the inflaton field, but when a tachyonic instability of the waterfall field triggers its end. 

Here, we will briefly describe single-field inflationary models with a waterfall-like stage. Single-field models are simpler to analyze, and the results confirm the increase in $n_s$ found in the hybrid inflation scenario \cite{Kallosh:2022ggf} (which moves in the direction preferred by current datasets). We will show here the example of the single-field waterfall-modulated inflationary models \cite{Zhang:2026ivx,Yuan:2026xcg,Renata1} which  can be used in various $\alpha$-attractor models, including quintessential inflation models that describe both inflation and dark energy.  But  we show in \cite{Renata1} that other functional forms of the  waterfall mechanism  increasing $n_s$ are also available.

We present preliminary results of our work in progress \cite{Renata1,Renata2}, stimulated by the studies of various datasets and their constraints on inflationary models in \S\ref{sec: results}. One example of a waterfall modulated T-model $\a$-attractor potential  is 
\be
V=  V_{0 }\tanh^2\Big({ \vp\over \sqrt{ 6\a}} \Big) \Big  ( 1+\gamma \tanh\left ({\varphi-\varphi_c\over \Delta \varphi}\right)\Big) \ .
\label{exampleT}\ee
Here the potential has a waterfall at $\vp=\vp_c$  with its size and gradient controlled by $\gamma$ and 
$\Delta \vp$, respectively. At $\gamma=0$ it is a normal $\a$-attractor T-model.

\begin{figure}[h]
\vskip -0.5cm 
\centering
		 \includegraphics[width=0.66\textwidth]{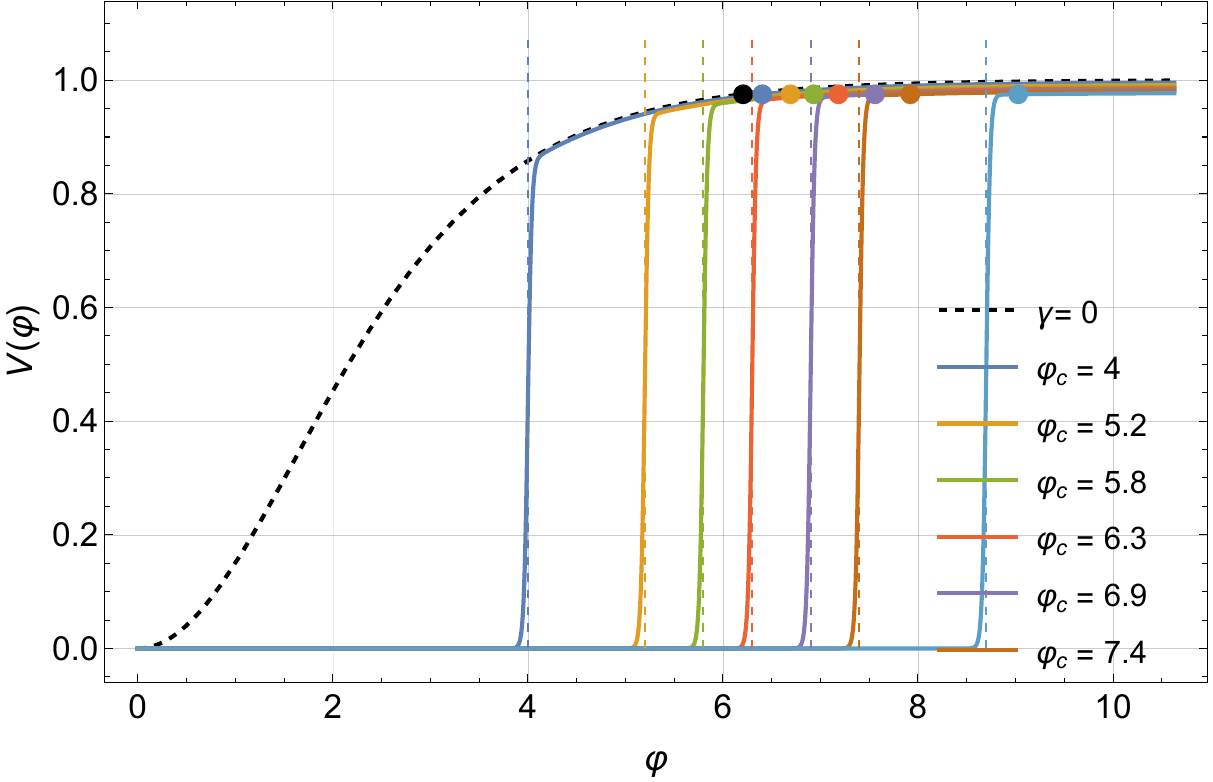}
        \includegraphics[width=0.7\textwidth]{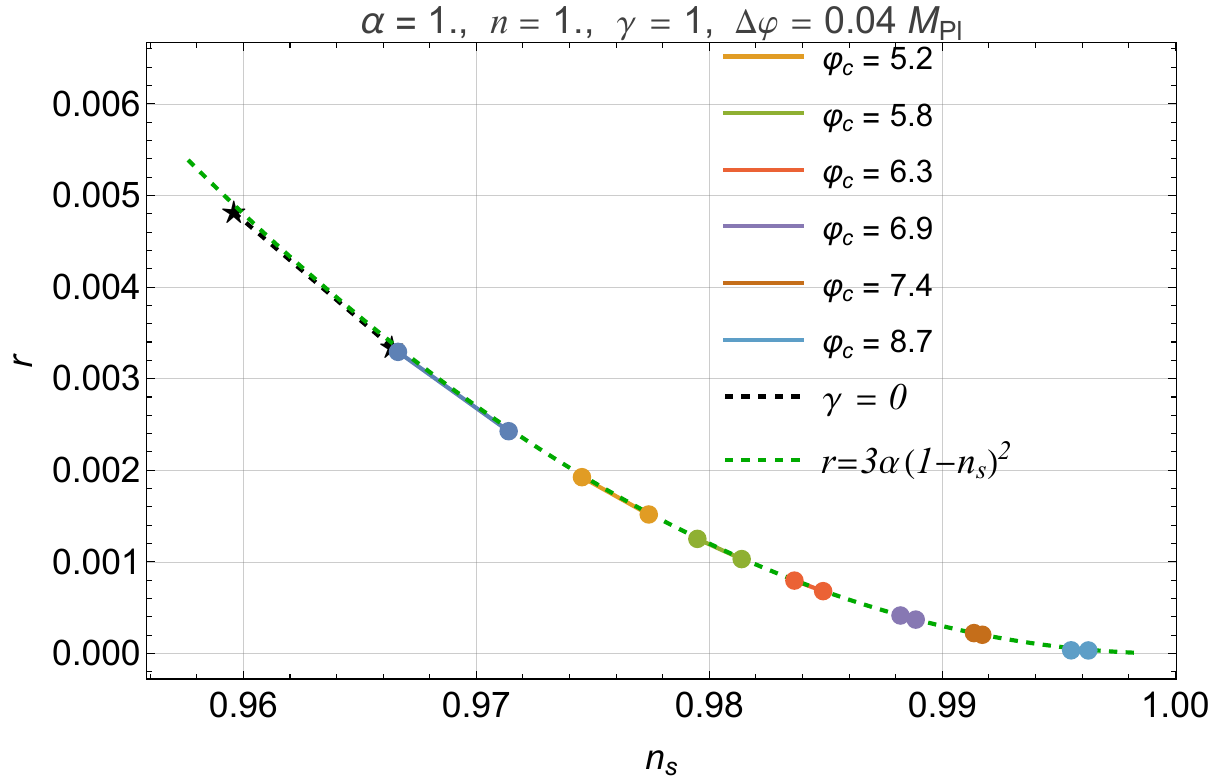}
        \caption{\footnotesize \textit{Upper panel}: potentials of the $\a=1$ waterfall-modulated T-models, with the no-waterfall equivalent shown by a black dashed curve. All other potentials rapidly descend due to waterfalls at various values of $\vp_{c}$.  Colored markers are the points where the last 50 or 60 e-folds of inflation begin for each choice of $\vp_{c}$.  Inflation is almost instantly terminated at $\vp \approx \vp_c$. 
        \textit{Lower panel}: $(n_s, r)$ values for the models corresponding to the potentials shown in the upper panel. The colored markers here show the values of the $(n_s, r)$ for  50 to 60 e-foldings, for each $\vp_c$. On the upper panel, the positions of  $\vp_{c}$ for 50 or 60 e-folds are so close that they are shown by a single color marker, but the resulting values of $n_{s}$ for 50 or 60 e-folds are distinctly different, as shown in the lower panel.
Changing $\vp_c$, the position of the waterfall, one can densely populate the entire curve $r\approx 3\a(1-n_s)^2$ in the lower panel.}
\label{TandPot}
\end{figure}

 The step potentials of the kind $V=V_{\rm original} \cdot V_{\rm step}$ where, $V_{\rm step}= 1+\gamma \tanh\left ({\varphi-\varphi_c\over \Delta \varphi}\right)$ were introduced in \cite{Adams:2001vc} and have been used in a number of cosmological applications. In particular, in \cite{Zhang:2026ivx,Yuan:2026xcg,Renata1,Renata2} this expression was used in the context of single-field inflationary step-modulated models, and in \cite{Renata1}, we have applied it to waterfall-modulated $\alpha$-attractors. We present here in Fig.~\ref{TandPot} an example of a waterfall-modulated T-model with a nearly instantaneous end of inflation due to very steep waterfalls.
The basic idea of this scenario is especially simple to explain in the case of $\gamma = 1$ and $\Delta \vp \ll 1$. In this scenario, we have
\be
V=  2 V_{0}\,\tanh^2\Big({ \vp\over \sqrt{ 6\a}} \Big)  \quad    \rm{for}  \quad \vp - \vp_{c} \gg \Delta\vp, \qquad  V=0  \quad {\rm for}  \quad \vp_{c} - \vp \gg \Delta\vp  .
\label{exampleT2}
\ee
In this scenario, inflation occurs only for $\vp \geq \vp_{c}$.  All derivatives of the potential are exponentially suppressed by $e^{-{2\over 3\alpha} \vp} \lesssim e^{-{2\over 3\alpha} \vp_{c}} $. Therefore, for large $\vp_{c}$ the deviation of $n_{s} $ from $1$ becomes exponentially small, i.e. $n_{s}$ is extremely close to 1.  But if we gradually decrease $\vp_{c}$, the derivatives of the potential gradually grow, and the values of $n_{s}$ gradually decrease until they reach their standard $\alpha$-attractor value $n_{s } = 1-2/N$. This scenario is illustrated in Fig.~\ref{TandPot}, and a more detailed discussion can be found in \cite{Kallosh:2022ggf,Renata1}.   

The slow-roll parameters $(r, n_s)$ in simple $ \alpha$-attractor models \cite{Kallosh:2013yoa} depend on $ \alpha$ and the number of e-foldings $N$, with $
n_s\approx 1-{2\over N}$, $ r\approx {12 \alpha\over N^2}
$ in the limit of large $N$.
This leads to the general relation \rf{nr} between $r$ and $n_s$ for all $N \gg 1$: 
\be\label{rel} 
 r(n_s) \approx  3\a(1-n_s)^2 .
\ee
In \cite{Renata1} we study various waterfall-modulated $\alpha$-attractor models 
where we employ premature termination of inflation via a waterfall insertion in single-field models.  An example of the instantaneous waterfall-modulated T-model is given here in Fig.~\ref{TandPot}. In general, we find that in these models the values of $( n_s, r)$  also satisfy the $r(n_s)$ relation in \rf{rel} for various choices of the height, width, and positions of the waterfalls.

\begin{figure}[h]
\vskip 0.5cm 
\centering
		 \includegraphics[width=0.8\textwidth]{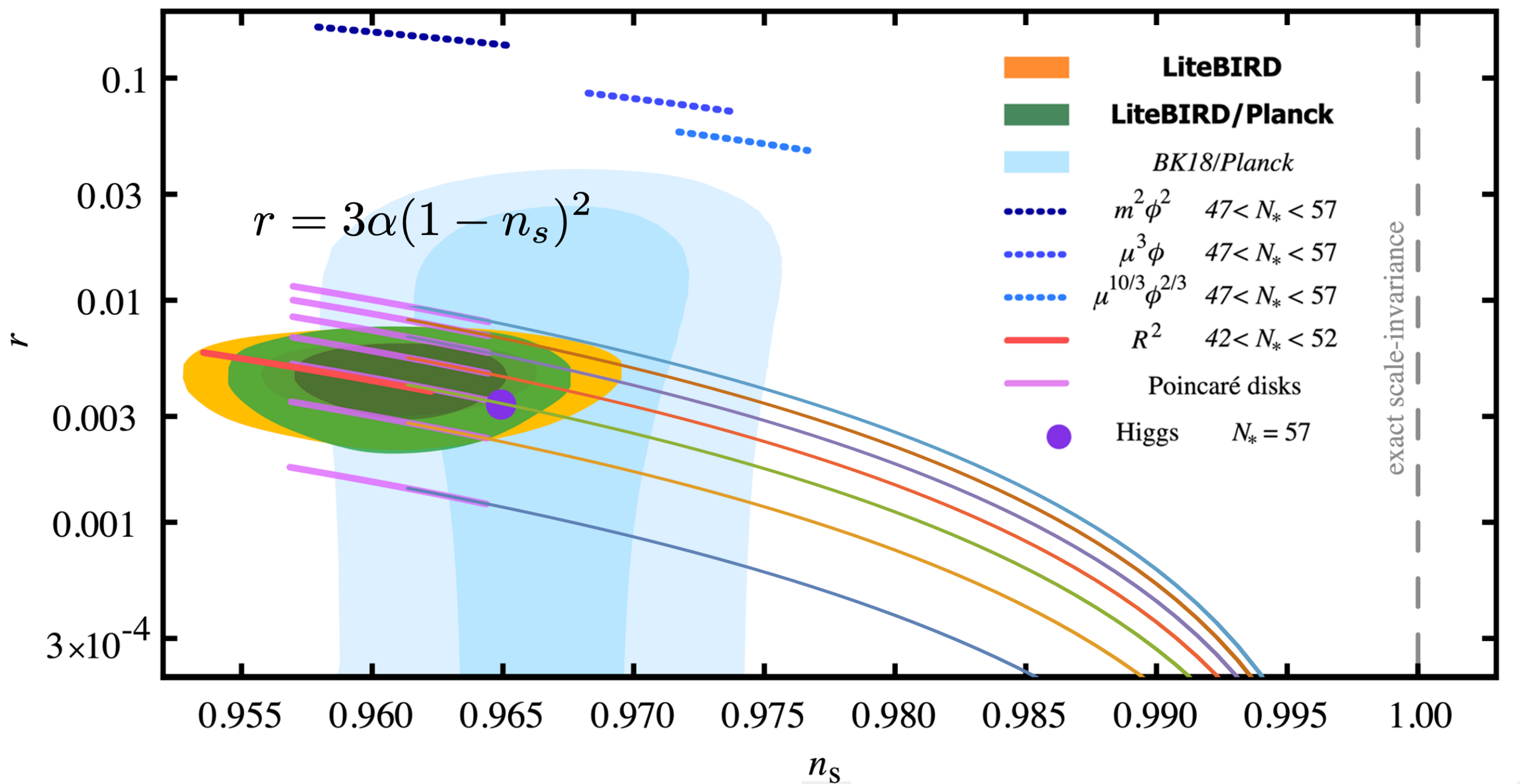}
        \caption{\footnotesize The seven descending horizontal curves show the forecast for $(r, n_s)$ for the waterfall models with increasing $n_s$ and decreasing $r$, along the curves  $r\approx  3\a(1-n_s)^2$. A discrete set of $\alpha$-attractors (with integer values of $3\a = 1, 2, 3,...,7$ given by purple lines) known as Poincar\'e disks  \cite{Kallosh:2013yoa,Ferrara:2016fwe} are present in the original LiteBIRD forecast \cite{LiteBIRD:2022cnt}.    In this figure, we extend these purple lines with the lines corresponding to models with a waterfall.}
        \label{Flux2}
\end{figure}
 
The above analysis yields targets for the LiteBIRD satellite from the waterfall-modulated $\alpha$-attractors. As shown in Fig.~\ref{Flux2}, the models are flexible in their value of $n_s$, generalizing the $n_s\lesssim 0.966$ targets obtained previously in the absence of waterfalls \cite{LiteBIRD:2022cnt}. The new targets from the waterfall-modulated $\alpha$-attractors stretch all the way to $n_s\approx 0.993$, with decreasing $r$, densely covering the curve $r=3(1-n_s)^2$.   The model illustrated in Fig.~\ref{TandPot} shows one of these 7 lines, the $\a=1$ case, the 3rd from the bottom Poincar\'e disk.

\subsection{\boldmath{$\a$-attractor  quintessence }}
The implications of dynamical dark energy for $n_s$ in the context of the phenomenological $\wa$ and $\owa$ scenarios, which can accommodate the observationally preferred evolution of the dark energy sector, are explored in \S\ref{sec: results}. The parameter constraints derived in the $\wa$ and $\owa$ models are presented in Tab.~\ref{tab:param}; see also the theoretical study of quintessence in an open universe \cite{Andriot:2026lac,Bhattacharya:2024hep}.

The simplest models of inflation and dynamical dark energy are based on a theory with two independent scalars: the inflaton, which drives inflation, and the quintessence field, which governs the evolution of dark energy.
The single-field theoretical quintessence models with a canonical kinetic term and scalar potential imply $w(z)\geq -1$ and therefore do not accommodate the observationally preferred evolution, in particular a crossing of the phantom divide, $w_{\rm DE}=-1$, favored by the combined CMB and DESI BAO data~\cite{DESI:2024mwx,DESI:2025zgx}.

A variety of different solutions to this problem have been suggested. Here we will discuss a two-field  quintessence model proposed in \cite{Toomey:2025yuy},  which shows that a two-field  quintessence model can imitate an effective phantom phase in the dark-
energy equation of state, whilst the total system always remains non-phantom and satisfies the Null Energy Condition. 

Our observation here is that the model in \cite{Toomey:2025yuy} is 
 actually  an $\a$-attractor model based on a hyperbolic geometry. The \K curvature of the hyperbolic geometry is 
\be
\mathcal{R}= - {2\over 3\a}
\ee
The  two-field  quintessence models proposed in \cite{Toomey:2025yuy} depends on a complex field,   the ``half-plane coordinate''  of the hyperbolic geometry
\be
T= e^{\pm \sqrt{2\over 3 \a} \vp} + i \sqrt{2\over 3 \a}\,  a
\ee
This complex field  represents a dilaton-axion pair $(\vp, a)$ via its real and imaginary parts. This is analogous to \cite{Linde:2018hmx}, which studied a  dilaton-axion pair $(\vp, a)$ with the same hyperbolic geometry kinetic term, but a different potential, with the purpose to produce  primordial black holes and gravitational waves.

 The axion-dilaton quintessence action  in \cite{Toomey:2025yuy} as a function of the complex field $T$ is\footnote{We replace the $\a$ in eq. (5) in \cite{Toomey:2025yuy} by $\beta$ to avoid confusion with $\a$ defining the curvature of the moduli space in these models.}
\be
{1\over \sqrt{g}} \, {\cal L} (T, \bar T)= -3\a {\partial T \partial \bar T\over (T+\bar T)^2}- V_0  \Big ({T+\bar T\over 2} \Big)^{\mp{\beta\over \sqrt{| \mathcal{R}|} }} 
 - m^2 \, f_a^2 \Big [1-\cos{T-\bar T  \over  2 i   \sqrt{| \mathcal{R}|} f_a}\Big ]
\label{action}\ee
A kinetic term in  the hyperbolic geometry with $SL(2, \mathbb{R})$-symmetry is
\be
-3\a {\partial T \partial \bar T\over (T+\bar T)^2}= -{1\over 2}[(\partial \vp )^2 + e^{\mp 2\sqrt{2\over 3 \a}\vp}  (\partial a )^2]
\ee

Comparing this with the kinetic term in \cite{Toomey:2025yuy}, we find that the parameter $\lambda$ in \cite{Toomey:2025yuy} is
\be
\lambda= \mp 2\sqrt{2\over 3 \a}
\ee
The potential $V(T, \bar T)$ in \eqref{action} breaks $SL(2, \mathbb{R})$-symmetry, in particular, it breaks the axion shift symmetry. One can check that the potential in \eqref{action} is 
\be
V(T, \bar T)\to  V(\vp, a) = V_0 e^{-\beta \vp} + m^2 \, f_a^2 [1-\cos(a/f_a)]
\label{Mpot}\ee
Thus, both kinetic and potentials  in the two-field quintessence model in \cite{Toomey:2025yuy}, up to change of notations can be represented using hyperbolic geometry where $\lambda= -2\sqrt{2\over 3 \a}$ and $\beta $ is an independent parameter. 

An even more interesting situation can be obtained by embedding the two-field quintessence model  \cite{Toomey:2025yuy} into the quintessential model $\a$-attractor model Exp II described in \cite{Dimopoulos:2017zvq,Akrami:2017cir,Zhumabek:2023wka,Jing:2026ymp}. This model describes both inflation and dark energy for a single field $\vp$. At the dark energy stage it already has a term of the form $V_0 e^{-\beta \vp}$ where 
\be
\beta = \sqrt{2\over 3 \a}= {|\lambda|\over 2}
\ee
In this model we have to add the axion term to potential  to  build a  consistent  axion-inflaton  quintessential $\a$-attractor model. We will  study this model  in \cite{Renata2}.

\begin{figure}[H]
\vskip 0.3cm 
\centering
		 \includegraphics[width=0.6\textwidth]{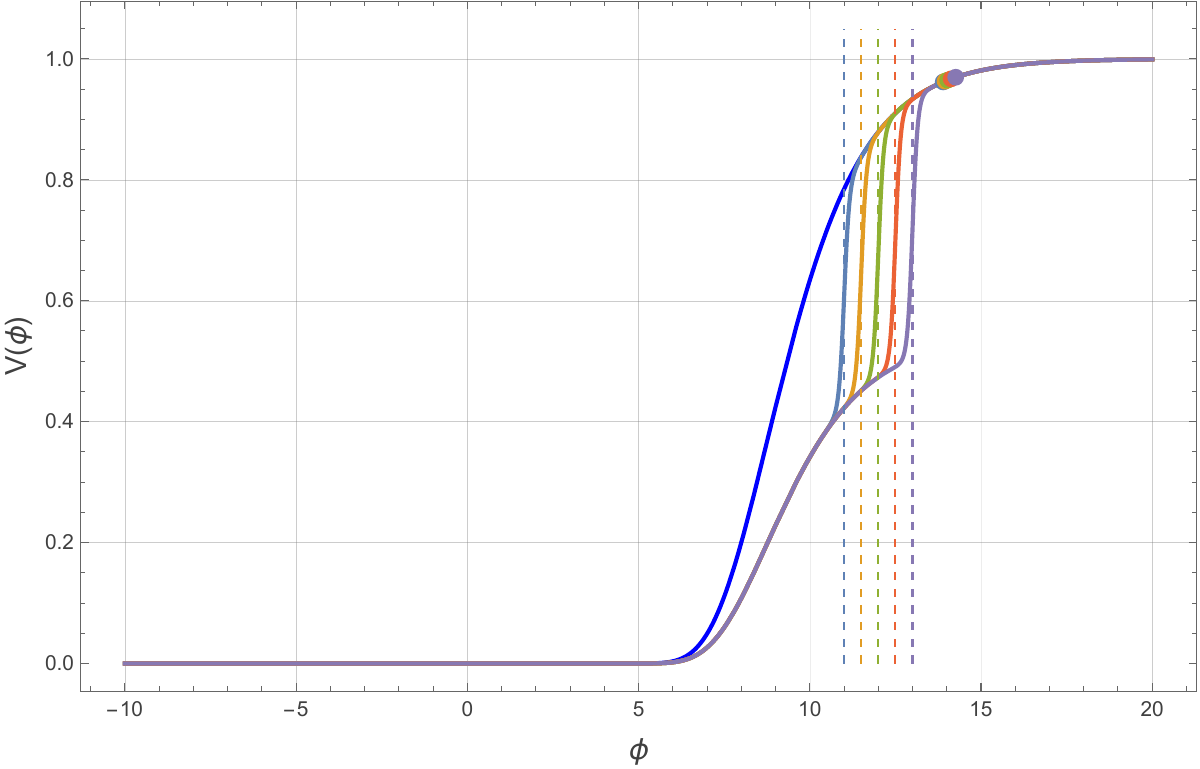}
		\caption{\footnotesize The blue curve describes the original quintessential two-plateau potential, studied in \cite{Dimopoulos:2017zvq,Akrami:2017cir,Zhumabek:2023wka,Jing:2026ymp}. The upper plateau represents the inflationary plateau; the lower plateau represents the dark energy stage; in between, there is a kination stage. The five waterfall-modulated   curves show  updated quintessential $\alpha$-attractor  potentials which prematurely terminate inflation and allow a gradual increase of $n_s$ from 0.964 up to 0.975 for $N=55$.}
        \label{Q}
\end{figure}

We further note that for the two-plateau quintessential $\a$-attractor model Exp II \cite{Dimopoulos:2017zvq,Akrami:2017cir,Zhumabek:2023wka,Jing:2026ymp} we can employ the same mechanism of premature termination of inflation via a waterfall insertion. In this updated quintessential $\alpha$-attractor model, we are able to increase $n_s$ gradually by adjusting the waterfall features in the new models. The relevant potentials are shown in Fig.~\ref{Q}. 

Future experiments will be able to test this model and its generalizations by obtaining tight constraints on both $(n_s, r)$ and $w(z)$, and elucidating whether the late-time expansion history is noticeably different from that of a cosmological constant.

\section{Discussion}\label{sec: discussion}

As discussed in the Introduction, exponential plateau inflationary models, such as the Starobinsky, Higgs, and simplest exponential $\alpha$-attractor models, are somewhat disfavored by the CMB data when combined with the DESI DR2 data. However, these conclusions should be interpreted with caution due to the tension between CMB and DESI data within the $\Lambda$CDM framework.

In this work, we have re-evaluated the above issues, allowing for certain deviations from the $\Lambda$CDM paradigm. In particular, we found that CMB + DESI data favor a slightly open Universe with $\Omega_k\simeq 3\times 10^{-3}$ at the $2.7\sigma$ level, which lowers the inferred value of the spectral tilt $n_s$ by $1.8\sigma$ relative to the $\ld$ result. Consequently, the tension with the predictions of the Starobinsky and  Higgs models, as well as the T-models of exponential $\a$-attractor inflationary scenarios, is reduced to statistically acceptable levels of $2.3\sigma$ and $1.5\sigma$, respectively. The tension with E-models of $\a$-attractors may be even smaller since they predict somewhat higher values of $n_{s}$ than T-models \cite{Kallosh:2025ijd}.

From the point of view of cosmological observations, any values of $\Omega_k$ represent possible cosmological models, independently of their sign. That said, it is not easy to deviate from $\Omega = 1$ in inflationary theory, and it may be even more challenging to explain the smallness of deviations from $\Omega = 1$. 

The simplest mechanism of creating an open universe in inflationary cosmology is based on tunneling and formation of bubbles representing open universes. As mentioned in \S\ref{sec: theory}, one can do it by tunneling through a sharp peak of the inflationary potential \cite{Linde:1998iw,Sasaki:1998ug, Linde:1999wv, Yamauchi:2009zz, Yamauchi:2011qq,Freivogel:2005vv,Bousso:2013uia}. In the simplest inflationary scenarios, including the Starobinsky model, Higgs inflation, or simple $\alpha$-attractor models, such peaks do not arise naturally. One can add such peaks, but this would diminish the beauty of these models.  Of course, one may simply delegate the task of creating an open universe to phase transitions between various vacua in the landscape, but such transitions may interfere with inflation in the models discussed in this paper. An intriguing alternative possibility is that we live in a compact, topologically non-trivial open universe, which could lead to a specific, potentially observable CMB anisotropy \cite{Jones:2023ncn}.  In either case, one would need to explain why $\Omega_k\simeq 3\times 10^{-3}$ is so close to zero but not zero, perhaps with anthropic explanations \cite{Freivogel:2005vv}.
 
As seen above, inflation does not provide a strong argument for an open universe. That said, it does not rule it out either. In general, inflationary models predicting $\Omega = 1$ are much more natural than those with non-zero curvature. The possibility of slightly changing the shape of the potential to match the recent CMB+DESI dataset, which has been discussed in many recent papers (e.g., a switch from exponential to polynomial attractors \cite{Kallosh:2025ijd}) is much easier to achieve than to explain the origin of sharp peaks in the slow-roll potential required in the open bubble universe scenario, or to speculate about an interplay between the swampland constraints and anthropic considerations in the universe with a non-trivial topology.
However, the recent developments in observational cosmology are truly spectacular. Thus, it is important to be aware of various ways to modify some features of the cosmological theory without sacrificing its basic principles. This may be helpful if observational data indicating a marginally open universe \cite{DESI:2025zgx,SPT-3G:2025bzu,Chen:2025mlf,Chudaykin:2025lww} or tiny deviations from the universe's isotropy \cite{Jones:2023ncn} continue to accumulate.

We also find that allowing for time-evolving dark energy improves the agreement with the exponential plateau models, although the effect is more modest than that of spatial curvature.  One cannot do it in the Starobinsky model or Higgs inflation without adding a new field, quintessence.  Meanwhile $\alpha$-attractors can describe inflation and quintessence by a single field \cite{Dimopoulos:2017zvq,Akrami:2017cir,Zhumabek:2023wka,Jing:2026ymp}. In \S\ref{sec: theoryDE}, we discussed new inflationary models \cite{Renata1,Renata2} with flexible values of $n_s$ that can appear in models with and without dynamical dark energy. A short summary of these models is presented in Fig.~\ref{Flux2}. It shows an extended forecast for LiteBIRD, which was originally based on exponential $\a$-attractors \cite{Kallosh:2013yoa,Ferrara:2016fwe} and given by the seven Poincar\'e  disks in   Fig.~2 of \cite{LiteBIRD:2022cnt}. These are purple lines at $n_s\lesssim 0.965$. The new models described in \S\ref{sec: theoryDE} shift the forecasts of the seven Poincar\'e  disks downward and to the right as $n_s$ increases and $r$ decreases, following the consistency relation $r=3(1-n_s)^2$.

While we have found that spatial curvature (and to an extent dynamical dark energy) yields an improved fit to the combined cosmological datasets and theoretically favorable values of $n_s$, it is important to note that this is not the only solution to the current tensions plaguing observational cosmology. As shown in a number of works, the preference of CMB-plus-DESI for high $n_s$ is broadly sourced by differences in the low-redshift reconstructed expansion histories and thus the $\ld$-derived values of the matter density parameter, as well as the sound-horizon parameter $r_dh$ \citep[e.g.,][]{Ferreira:2025lrd,Zaldarriaga:2026itn,Loverde:2024nfi,DESI:2025zgx}. This can be interpreted as evidence for dynamical dark energy \citep[e.g.,][]{DESI:2025fii}, curvature \citep[e.g.,][]{Chen:2025mlf}, small-scale topology \citep{Philcox:2025faf}, modified recombination histories \citep{Sailer:2025lxj}, negative effective neutrino masses \citep[e.g.,][]{Loverde:2024nfi,Graham:2025dqn,Green:2024xbb}, and beyond. Indeed, if the differences turn out to be a statistical or systematic fluctuation, which resolves to find consistent $\Omega_m$ and $r_dh$, one expects much of the $n_s$-discrepancy to disappear. Future data from both the CMB and galaxy surveys will shed light on this issue and decisively demonstrate whether the current preferences for an open universe and/or dynamical dark energy are physical in origin.

\textbf{Note added.}
When this paper was in the final stage of preparation, an analysis of non-minimal cosmological models appeared~\cite{Giare:2026oti}, which includes discussion of the spectral index. The results presented in \S\ref{sec: results} are in good agreement with \cite{Giare:2026oti}; however, our analysis additionally includes the DESI DR1 full-shape galaxy clustering data (the redshift-space power spectrum and bispectrum monopole). These datasets have a non-trivial impact on the constraints on $n_s$ and therefore on the conclusions regarding inflationary scenarios.


\section*{Acknowledgements}
\footnotesize 
AC acknowledges funding from the Swiss National Science Foundation. RK and AL are supported by Leinweber Institute for Theoretical Physics at Stanford and by the NSF grant PHY-2310429. RK and AL are grateful to Y. Akrami, G. Alestas and M. Shmakova for the discussions of dark energy in the context of quintessential $\alpha$-attractors. 
YY is supported by IBS under the project code IBS-R018-Y3-2026-a00. YY would like to thank Swagat Saurav Mishra for useful discussion and detailed explanation about the spatial curvature on the BAO data.

\bibliographystyle{JHEP}
\bibliography{lindekalloshrefs}
\end{document}